# 基于深度学习文本情绪挖掘股市相关性研究

张宸瑞
(华南理工大学 经济与金融学院,广州 510006)

**摘 要**:探讨如何对股吧等金融论坛数据进行爬取并结合深度学习模型进行情感分析,本文将使用 BERT 模型针对金融语料进行训练,并对深证成指进行预测,通过最大互信息系数对比研究发现将 BERT 模型应用到金融语料中,所得到的情感特征能够反映在股票市场的波动,有利于有效提高预测准确度。同时本文将深度学习与金融文本相结合,在通过深度学习的方法进一步探究投资者情绪对股票市场的影响机制,将有利于国家监管部门和政策部门对维持股票市场稳定性制定更加合理的政策方针。

**关键词**:情感分析;深度学习;BERT;最大互信息系数;

# Research on the correlation between text emotion mining and stock market based on deep learning

ZHANG Chenrui
(School of Economics and Finance, SCUT, Guangzhou 510006, China)

**Abstract:** This paper discusses how to crawl the data of financial forums such as stock bar, and conduct emotional analysis combined with the in-depth learning model. This paper will use the Bert model to train the financial corpus and predict the Shenzhen stock index. Through the comparative study of the maximal information coefficient (MIC), it is found that the emotional characteristics obtained by applying the BERT model to the financial corpus can be reflected in the fluctuation of the stock market, which is conducive to effectively improve the prediction accuracy. At the same time, this paper combines in-depth learning with financial texts to further explore the impact mechanism of investor sentiment on the stock market through in-depth learning, which will help the national regulatory authorities and policy departments to formulate more reasonable policies and guidelines for maintaining the stability of the stock market.

**Key words:** emotion analysis; deep learning; BERT model;Maximal information coefficient (MIC)

# 一、问题的提出

股市的交易实际上是人与人的博弈，而这种博弈也是不少学者正在研究的方向，在金融领域中，股票价格的波动也牵动着大家的心，20世纪70年代，美国经济学家尤金·法玛提出了"有效市场假说"，假说认为在有效市场中，任何时刻的股票价格都充分准确地反映了全部市场信息。在20世纪80年代后，随着市场中大量市场异像和非理性的交易行为的出现，"有效市场假说"受到了很大的挑战。据统计A股股民数量已达到1.89亿，这是一个非常惊人的数字，其中自然人（散户）数量达到82%（数据来源：2018上海证券交易所），个人散户在投资领域的地位十分庞大。

根据贺行知（2021）[1]的结论，从我国股市发展阶段来看，我国股票市场已初步具备弱有效性市场的基本特征，但是，股市存在投资者利用基本面消息获取超额利润的现象，股市并非半强有效市场。John Maynard Keynes（1936）曾指出许多经济活动都受"动物精神"的支配，行为金融也依此营运而生。Shiller(2017)指出，"随着研究方法的推进，随着社交媒体数据积累的增加，文本分析将成为未来几年经济学领域更强有力的领域。我们的目标是推动研究朝着这个方向发展。"近年来，研究投资者情绪与资产定价间的作用关系逐渐成为热点，研究主要包括数据的选择、情绪的提取分析以及模型的构建。

# 二、文献综述与研究方法评述

## (一) 数据的选取

早期的研究主要选取结构化的数据来间接地反映投资者情绪，如市场信息（开盘价、收盘价、市场成交量等）、经济指标（宏观经济指标、财务报表数据）和技术指标（简单日移动平均线、加权日移动平均线等）随着近年来文本挖掘和数据分析技术的提升，越来越多的研究开始采用可直接反映投资者情绪的文本类非结构化数据，如新闻（财经新闻、普通新闻、公司报道等）、社交媒体（Twitter、Facebook、微信（石善冲，2018）[2]、微博、博客、股吧论坛等）。由于数据的易提取性和预测结果的准确性，使用股市数据和技术指标的研究还是占据主流，但是利用社交媒体数据进行研究的趋势在渐渐兴起。未来，将社交媒体数据与股市数据及指标相结合将会是一个很好的研究方向(Bustos,2020)。

## (二) 情绪提取与分析

对于结构化数据，首先是运用市场指标来代理人气，如交易量、封闭式基金折扣、首次公开发行（IPO）当日回报等。可以说，最具影响力的衡量标准是（Baker & Wurgler（2006））的投资者情绪指数，该指数是六个基于市场的代理的主要组成部分。第二种方法是以调查为基础。热门的消费指数包括密歇根大学消费者情绪指数和瑞银/盖洛普投资者乐观指数。对于非结构化的文本数据，则主要通过基于词典与规则的文字包技术

（TF-IDF）和机器学习方法（如 word2vec（Mikolov，2013））[3]从文本中提取出投资者情绪，并以积极、中立、消极等方式进行分类。目前，在学术界还没有公认的投资者情绪衡量指标，如何更好地提取和衡量投资者情绪还仍需不断探索（唐国豪，2016）[4]。

### (三) 情绪预测模型

近些年来，机器学习算法越来越多地应用到预测模型中，如支持向量机（SVM）、人工神经网络（ANN）、贝叶斯模型以及深度学习（主要有 CNN、ELM、LSTM、DBN 等方法）日益普及。深度学习是人工神经网络的一个子集，但不同于传统的机器学习算法，它不需要对数据进行预处理和提取特征。（Kraus & Feuerriegel 2017）研究发现深度学习算法对传统的神经网络算法在股市预测上有更高的准确度。利用机器学习的研究中，支持向量机方法仍被广泛应用，但深度学习算法的应用在未来很长一段时间内会是股市预测的研究热点（Bustos，2020）。

## 三、数据的选取

### (一) 文本数据的选取

本文通过 Python 爬取东方财富网 [1]股吧（zssz399001）2019 年 1 月 1 日至 2020 年 12 月 31 日的股吧数据，我们横向对比了个股以及指数股吧数据，最终选择了深证成指的股吧，相比上证综指以及其他指数股吧数据，其数据量更充足，浏览人数更多更能代表绝大多数股民的想法，其次，深证成指包括深圳证券交易所上市的具有一定规模性、优质的 500 家上市公司的股票。深证成指的编制也是抽取了各个行业板块的股票，因此深证成指能够很好的反映出深市的股票情况。在数据的处理上，本文遵循文本分析的标准，删除重复的数据，删除非文本项目如编码图像、表格、 HTML 标签和表情符号等。

---

[1] 根据 iResearch 2018 年的一份分析报告指出，东方财富网是中国顶级金融网站。每月有效浏览时间为 7800 万小时，占市场份额的 45%，高于前 10 家公司中其余 9 家公司的总和。

## 表 1 每日评论统计汇总信息

| 时间 | 评论 | 阅读量 | 评论数 | 网页来源 |
|---|---|---|---|---|
| 2020-9-9 | 今天抄底爽歪歪了 | 95 | 0 | http://guba.eastmoney.com/news,zssz399001,9634 |
| 2020-9-9 | 创业板公司亚光科技：股东合计减持25%的股份，这不是在减持，分明是在找人接盘准备 | 793 | 1 | http://guba.eastmoney.com/news,zssh000001,963406436.html |
| 2020-9-9 | 如果我专心玩白银可能都不会输这么惨，去年账户上的8万现在只有2万了。 | 317 | 3 | http://guba.eastmoney.com/news,zssz399001,963400740.html |
| 2020-9-9 | 美哥跌耶，我也跌耶，我比美哥跌得黑耶 | 301 | 3 | http://guba.eastmoney.com/news,zssz399001,963404635.html |
| 2020-9-9 | 炒小、差是投资者最基本的选择权利，关键是违规违法没有，既然规则已制定，如果没有违则涨跌应该交给市场 | 183 | 0 | http://guba.eastmoney.com/news,zssz399001,963402753.html |

从表 1 所得到的统计信息可以看出，每天平均评论标题长度为 22.8 个字符，从表中可以看出，内容呈右偏分布。与传统的（相对较短的）评论不同的是，股吧中数据一些很长的消息经常被从其他来源复制和粘贴，如新闻报道和分析报告。我们使用简单的处理来消除这些潜在的潜在影响离群值，仅保留少于 150 个汉字的消息。此外考虑到每日评论数参差不齐以及降低不同阅读量对结果的影响，我们选择了按阅读量排序，由高到低每日选取了前 50 条数据进行研究。

## 表 2 经预处理的股吧评论文本格式

|  | Mean | S.D | Skewness | Min | Max | Count |
|---|---|---|---|---|---|---|
| 数据长度（字符） | 22.86383 | 13.40368 | 0.126406 | 2 | 66 | 32298 |

注：此表为每天评论长度的汇总统计信息。每个变量的样本均值、标准差（S.D.）、最大最小数和评论总数。该样本包含 2019 年 1 月 1 日至 2020 年 12 月 31 日样本期内深证成指股吧评论数据。

## (二) 股票交易数据的选取

本文选取深证成指作为研究对象，其中选取每日交易数据进行研究，包括当天的收盘价、开盘价、最高价、最低价、昨收价、涨跌额、涨跌幅、成交量、成交额等数据，数据的具体格式如表 3 所示。本文通过开源的 Python 数据 API——Tushare[2]获取，返回的结果为 Pandas.DataFrame 数据类型。

表 3 未经过归一化处理的行情数据

| 日期 | 收盘价 | 开盘价 | 最高价 | 最低价 | 昨收价 | 涨跌额 | 涨跌幅 | 成交量 | 成交额 |
| --- | --- | --- | --- | --- | --- | --- | --- | --- | --- |
| 20201231 | 14470.68 | 14226.28 | 14476.55 | 14226.28 | 14201.57 | 269.1178 | 1.895 | 3.72E+08 | 5.11E+08 |
| 20201230 | 14201.57 | 13970.45 | 14208.68 | 13968.09 | 13970.21 | 231.3549 | 1.6561 | 3.52E+08 | 4.69E+08 |
| 20201229 | 13970.21 | 14042.79 | 14082.5 | 13915.89 | 14044.1 | -73.89 | -0.5261 | 3.72E+08 | 4.78E+08 |
| 20201228 | 14044.1 | 14020.95 | 14112.59 | 13959.14 | 14017.06 | 27.0435 | 0.1929 | 3.73E+08 | 4.83E+08 |
| 20201225 | 14017.06 | 13879.24 | 14017.06 | 13835.52 | 13915.57 | 101.4832 | 0.7293 | 3.38E+08 | 4.35E+08 |

资料来源：Tushare, http://tushare.org/。

由于各种数据间的量纲可能不同，因此我们需要进行对数据进行归一化处理以保证数据在各个模型训练中保持一致的分布。归一化的方法我们使用离差标准化(Min-Max Normalization)(Patro 等，2015)[5]。主要是将特征映射到[0,1]之间，具体公式如下：

$$X = \frac{x - min(x)}{max(x) - min(x)} \quad (1)$$

表 4 离差标准化处理的行情数据

| 日期 | 收盘价 | 开盘价 | 最高价 | 最低价 | 昨收价 | 涨跌额 | 涨跌幅 | 成交量 | 成交额 |
| --- | --- | --- | --- | --- | --- | --- | --- | --- | --- |
| 20201231 | 1.00000 | 1.00000 | 1.00000 | 1.00000 | 1.00000 | 0.83034 | 0.73689 | 0.36816 | 0.44054 |
| 20201230 | 0.96354 | 0.96449 | 0.96310 | 0.96421 | 0.96747 | 0.80357 | 0.71986 | 0.33501 | 0.39241 |
| 20201229 | 0.93220 | 0.97453 | 0.94571 | 0.95698 | 0.97786 | 0.58719 | 0.56437 | 0.36716 | 0.40181 |
| 20201228 | 0.94221 | 0.97150 | 0.94986 | 0.96297 | 0.97406 | 0.65874 | 0.61560 | 0.36985 | 0.40789 |
| 20201225 | 0.93854 | 0.95183 | 0.93670 | 0.94584 | 0.95979 | 0.71151 | 0.65382 | 0.31269 | 0.35237 |

资料来源：Tushare, http://tushare.org/。

---

[2] Tushare 是一款基于 Python 语言的开源财经数据接口包,主要实现对股票等金融数据从数据采集、清洗加工到数据存储的过程,能够为金融分析人员提供快速、整洁和多样的便于分析的数据。

# 四、 对实体文本的情感分析

文本情绪感念提出（Bo Pang，2002）[6]后，早期对于文本情绪的度量主要在构建情感词典法，将文章中经常出现的表达情感的词进行赋分，并将其编纂成字典，通过使用字典对文章的匹配进行打分，这种方法通用性较强，游王靖一与黄益平在金融科技语境下构建情感词典,根据正负向情感词汇在文章中出现的频数、正负向情感词典中词的数量等指标分别对报道中的情感词赋予不同的权重,之后计算每篇报道中的正负情感指数,并通过直接加总获得报道的净情感指数[7]。

相较于情感词典法更加主观的赋分以及分类，机器学习法更加客观同时在不同领域的文本分析研究中有较好的表现,机器学习法的首要任务就是构造语料库。Al-Nasseri 与 Ali 将所选的公司在论坛中的与其公司有关的新闻文本资讯，使用朴素贝叶斯、决策树以及支持向量机（SVM）算法在软件中所训练的模型来预测[8]。Pawar 等人将递归神经网络（RNN）与长短期记忆单元（LSTM）相结合对股市进行预测并与传统的支持向量机、朴素贝叶斯分类器相比较[9]。

## (一)   模型设计

深度学习时代中的 NLP（自然语言处理）预训练工作广泛使用词嵌入（Word Embedding），使用深度学习模型进行训练的时候，会将所训练的次转化为词向量作为神经网络的输入层，而在深度学习模型训练的过程中，训练结果的好坏程度很大程度取决于训练集的大小，较大的训练集可以训练出较好的词向量，目前在自然语言处理领域绝大部分的任务模型都会采用训练好的词向量。在词向量的训练过程中，词向量忽略了上下文的表意，当词汇出现一词多义的情况，往往对应的是相同的词向量，这是不合理的，因此，2018 年 Devlin 等人提出预训练语言模型 BERT，一问世即刷新了 11 个 NLP 任务的榜单，是 NLP 领域前进的一大步[10]。BERT 模型的结构如图 1 所示，其中 $E_1,E_2,...,E_N$ 是模型的输入字符，输入字符通过双向的 Transformer 特征提取器获取文本特征，输入字符训练后输出相应的向量 $T_1,T_2,...,T_N$。

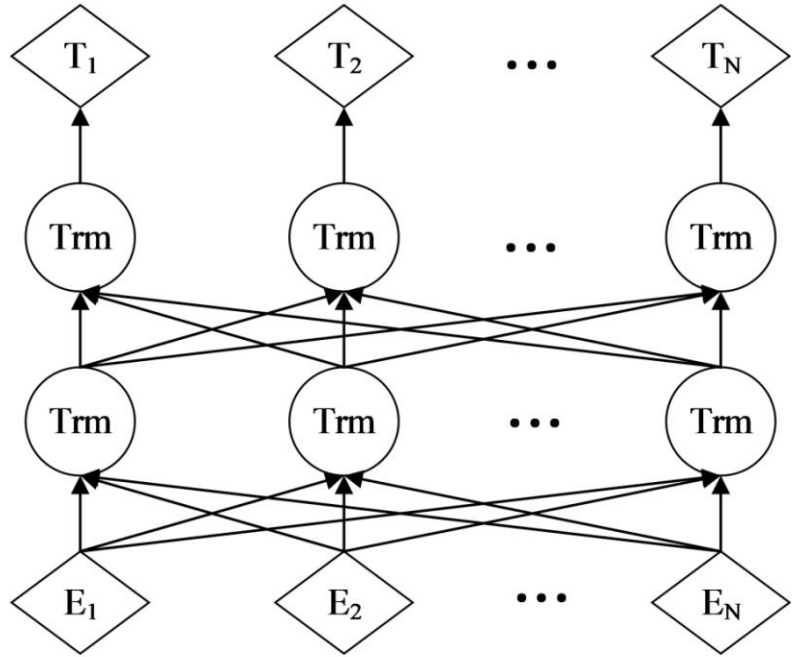

图 1 BERT 模型结构

如图 2，BERT 作为一个所有层都能结合上下文语义进行训练的模型，其输入由字嵌入（Token Embeddings）、段嵌入（Segment Embeddings）以及位置嵌入（Position Embeddings）三个向量组成，与此同时，BERT 采用的是 MLM 模型（遮盖语言模型），MLM 通过遮盖（Mask）一部分字，类似填空，然后去预测被遮盖的模型，通过迭代来达到上下文训练的目的。

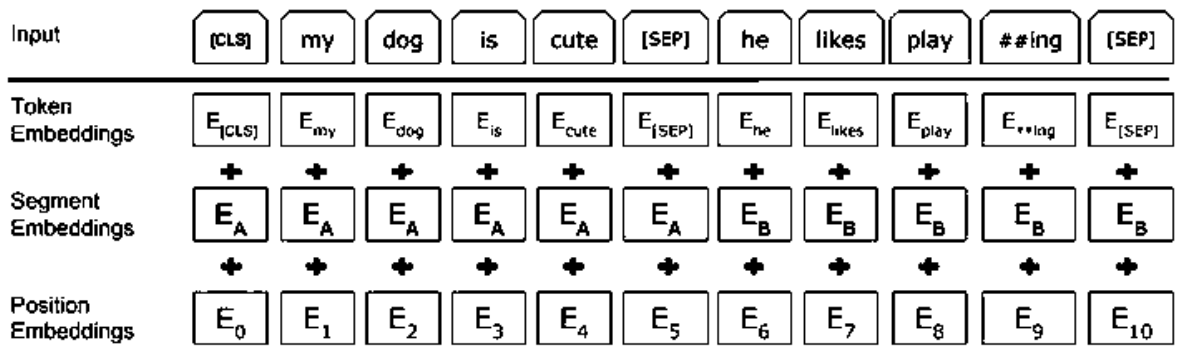

图 2 BERT 模型的训练方式

## (二) 基于金融语料的模型训练

BERT 本质上是一个两段式的 NLP 模型。第一个阶段叫做：Pre-training（预训练），利用现有无标记的语料训练一个语言模型，该阶段十分耗时，且对算力要求极高通常需要 4 到 16 个云 TPU 计算 4 天以上时间，由于受硬件限制以及对准确度的要求，我们选取哈工大讯飞联合实验室发布基于全词覆盖（Whole Word Masking，wwm）的中文 BERT 预训练模型 Chinese-BERT-wwm 作为预训练模型。第二个阶段叫做：Fine-tuning（微调），

利用预训练好的语言模型，完成具体的 NLP 下游任务。Pre-training 的训练成本很大，而 Fine-tuning 成本相对较少。我们正是在本地上，采取 Fine-tuning，使用金融语料对 BERT 模型进行训练。

语料在经过分词后输入 Encoder 模块得到转化后的索引，从而得到每一个词的词向量，这与 jieba 等分词工具的分词不同，例如："看来进入牛市了，大盘大涨，能带动投资情绪上涨"这句话分词之后会得到：[看，来，进，入，牛，市，了，大，盘，大，涨，能，带，动，投，资，情，绪，上，涨]，从而将词与 BERT 预训练模型中的语料表结合起来，在本文实验中，将最大词序列长度设置为 128 位，未满 128 位的将使用 0 进行填充，同时在句子的开头和结尾添加[CLS]与[SEP]标签。在 BERT 输入句子完成转化后，有两个训练方式，分别是 Masked LM 和 Next Sentence Prediction (NSP)下一句预测。

**1. Masked LM**

BERT 训练中在句子中使用[MASK]替换一部分词语，来使模型利用上下文进行预测，以"看来进入牛市了，大盘大涨，能带动投资情绪上涨"为例子，有 80%的概率将句子转变为"看来进入牛市了，大盘大[MASK]，能带动投资情绪上涨"，将句子中的涨用[MASK]代替，有 10%的概率保持句子不变，也有 10%的概率将"涨"用其他词代替例如："看来进入牛市了，大盘大跌，能带动投资情绪上涨"。这样 8:1:1 的替换策略主要是为了避免在后续的使用出现[MASK]的单词，从而导致性能受到影响。

**2. 下一句预测（NSP）**

BERT 训练中第二个任务为下一句预测，这样做的目的也是为了让模型在有监督学习下，能够结合上下文语义进行任务，同样以"看来进入牛市了，大盘大涨，能带动投资情绪上涨"为例子，在训练过程中，有 50%的概率将句子选择相连的两个句子："[CLS] 看来进入牛市了，大盘大涨，能带动投资情绪上涨[SEP]沪深两市翻红[SEP]"，同时也有 50%的概率选择不相关的句子连接："[CLS] 看来进入牛市了，大盘大涨，能带动投资情绪上涨[SEP]美股受大挫[SEP]"，同时在标签中输出"否"。

**3. 使用金融语料进行 Fine-tuning**

BERT 在完成预训练后，可将其用于金融实体情感识别的任务，在情感分析中[CLS]将作为下一网络的输出，根据金融文本的特殊性，使用带情感标注的金融文本进行 Fine-tuning，就可以训练出在金融等特定领域精度更高的模型。

# 五、 实验与分析

## (一) 实验环境

实验环境如表 5 所示。

表 5 实验环境

| 开发环境 | 参数 |
| --- | --- |
| 处理器 | R7-3750H(2.30GHz) |
| GPU | GTX1660Ti 6GB |
| 内存 | 16GB |
| 操作系统 | Windows 10 64 位 |
| 深度学习框架 | TensorFlow |
| 编程工具 | PyCharm |

## (二) 数据集

本文选用所爬取的 2019 年 1 月 1 日至 2019 年 12 月 30 日的数据，进行人工情绪的标注，本文使用三分类对情绪进行标注如表 6 所示，0 表示负向情绪、1 表示中性情绪、2 表示正向情绪，并对文本按 8:2 的比例划分训练集及测试集，进行训练。

表 6 情绪分类符号表示

| 负向 | 中性 | 正向 |
| --- | --- | --- |
| 0 | 1 | 2 |

## (三) 参数设定

本文模型选用工大讯飞联合实验室发布的 chinese_roberta_wwm_large_ext_L-24_H-1024_A-16（24-layer, 1024-hidden, 16-heads）预训练模型，即采用 24 层 Transformer，隐层维度为 1024，多头注意力的参数为 16，参数模型总大小为 330MB。模型训练方面批次大小（batch size）为 16，学习率（learning rate）为 2e-5，序列最大长度（max seq length）为 128。

## (四) 模型训练结果

训练结果为准确度为 0.7553，损失为 0.6558，如图 7 是预测结果的部分样例，如图 3 是根据训练的模型对 2020 年 1 月 1 日至 2020 年 12 月 31 日的评论数据的预测，可以看出负向的情绪居多，中性的情绪很少。

表 7 部分样例数据

| 时间 | 评论 | 情感 | 属于积极的概率 | 属于消极的概率 |
|---|---|---|---|---|
| 2020-6-9 | 北上今天净流入60亿，尾盘猛进二十亿，明天大盘无忧！ | 2 | 0.938666 | 0.0613335 |
| 2020-6-9 | 大跌正式开始 | 0 | 0.00298048 | 0.99702 |
| 2020-6-9 | 三家财务造假，暴风集团，东方金钰，长城影视，股价跌停 | 0 | 0.0382706 | 0.961729 |

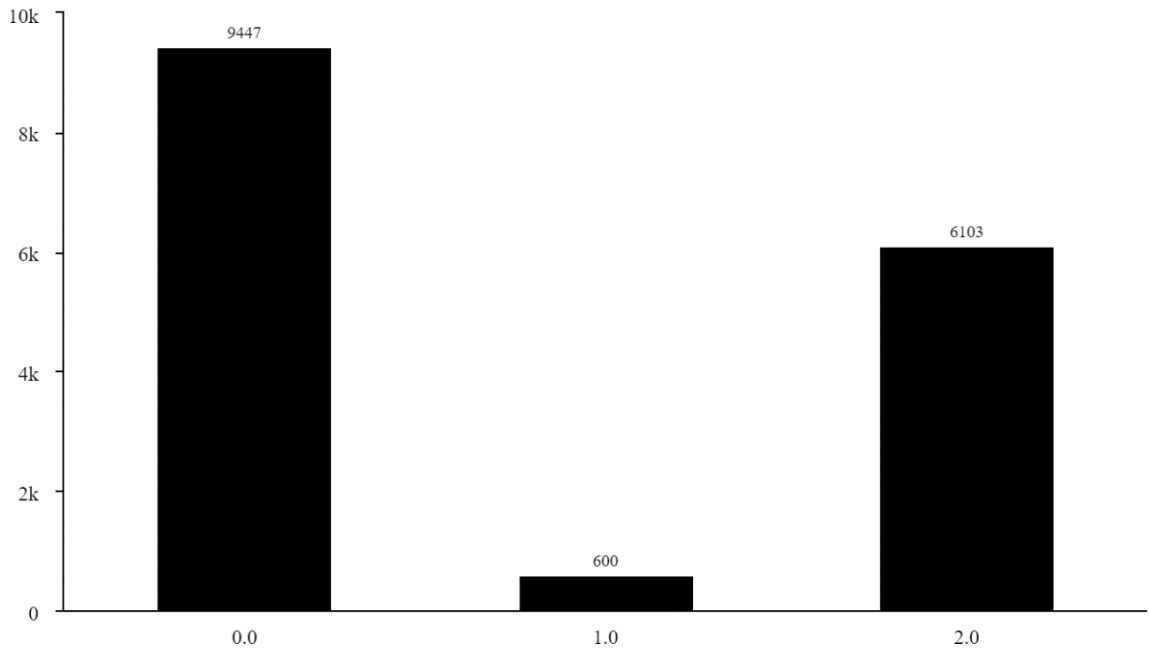

图 3 2020 年数据分布

## (五) 实证研究

### 1. 情绪数据的相关计算

对于每天有多条数据，本文使用对数据按日期分类进行处理的公式如下：

$$sentiment_t = P_{positive} + (-1) * P_{negative} \quad (2)$$

$$emotions^T = \frac{\sum_{t=1}^{n} sentiment_t}{n} \quad (3)$$

其中，sentimentt 表示每一条评论的情感得分，Ppositive、Pnegative 分别表示该条情绪为积极或消极的概率，情绪指数 emotionsT 表示在 T 日内所有情绪情感得分的平均数，emotionsT∈（0,1）。若 emotionsT 趋近于 0，则说明市场情绪消极，若 emotionsT，趋近于 1 则说明市场情绪积极。

同时由于各种数据间的量纲可能不同，因此我们需要进行对数据进行归一化处理以保证数据在各个模型训练中保持一致的分布，因此本实验数据使用公式 (1)进行归一化处理。

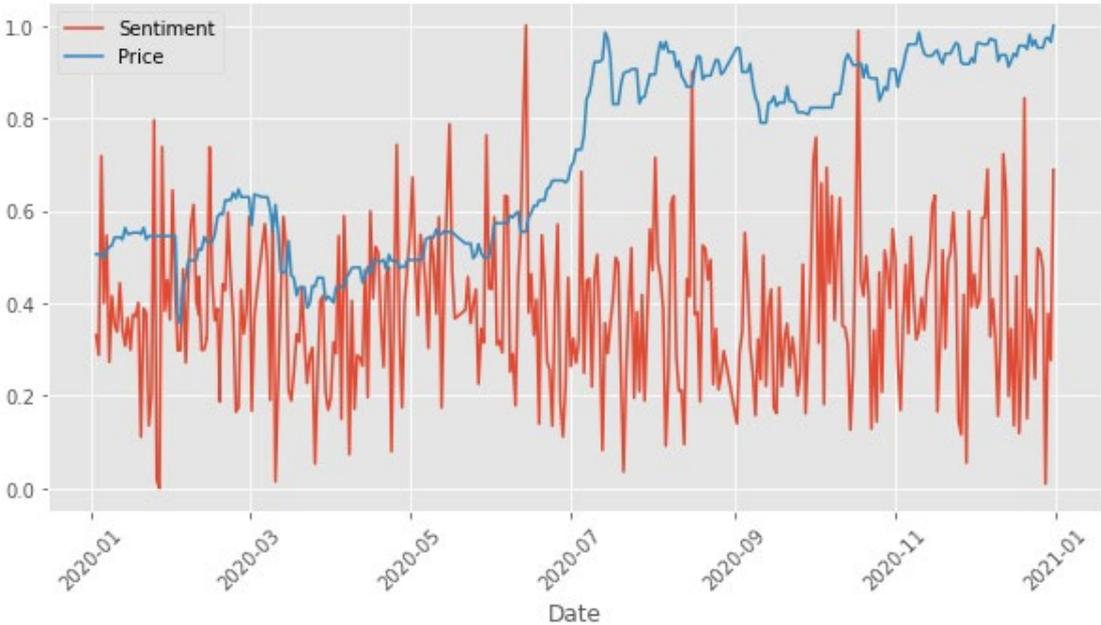

**图 4 市场情绪与股价归一化处理结果**

由图 4 可以看出，情绪指数与股价关系不是很明显，本实验按照时间窗大小为 30，每个窗口最少包含的观测值数量为 1 并对数据 Sentiment 和 Price 数据进行平滑处理，求出 Sentiment 和 Price 数据的 30 日均线，计算得出的 avg_Sentimen 和 avg_Price 代替原来的 Sentiment 和 Price，结果如图 5 所示，在市场情绪随着股票价格呈现同向波动。

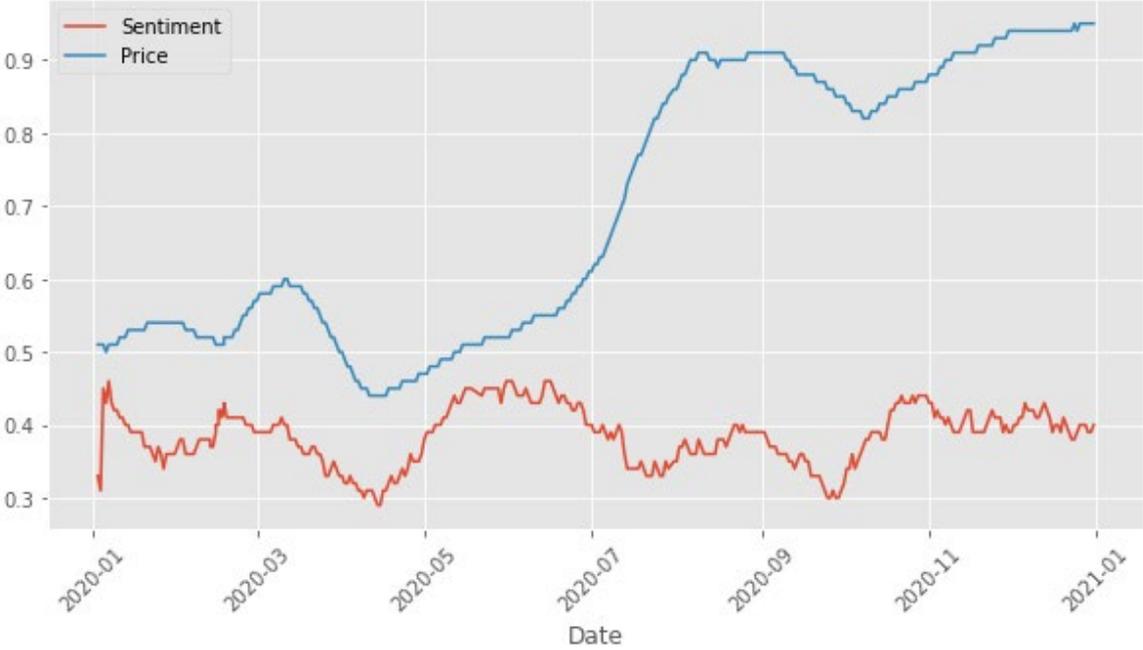

**图 5 市场情绪与股价平滑处理结果**

## 2. 基于最大互信息系数研究变量的相关性

### （1） 最大信息系数理论

2011 年哈佛大学的 David N. Reshef 等人[11]提出了最大信息系数(Maximal information coefficient)，简称 MIC, MIC 是衡量变量之间相互依存关系的一个很好的测度，它具有两个重要属性：广泛性和公平性。MIC 的广泛性是指它在多样本情况下对于多种函数关系都敏感，可以检测出多种关系类型，例如非函数关系和多种函数关系合成的超函数关系等。MIC 的均匀性是指当在不同的关系类型中加入相同的噪声时，它们之间的 MIC 值是相近的；反之，当计算出两个变量 MIC 值相似或者相等时，对于加入的噪声程度的值也相近。

### （2） MIC 的处理过程[12]

给定有限有序的数据集 X = {x1,x2,x3,…,xn}，如果将 x 轴划分为 x 个格子，y 轴划分为 y 个格子，那么就可以得到一个 x×y 的网格划分 G，其中 x，y 是正整数，将落入 G 的点的数量占 X 数量的比例看作是其概率密度 X|G，而根据不同的网格划分情况得到的概率分布 X|G 也不同。在 X = {x1,x2,x3,…,xn}中，两变量 xi 和 xj 之间的互信息可以定义

$$I(x_i, x_j) = \sum_{x_i \in X} \sum_{x_j \in X} p(x_i, x_j) \log_2 \left( \frac{p(x_i, x_j)}{p(x_i) p(x_j)} \right) \quad (4)$$

为：

在 x，y 给定的情况下，若改变 x，y 的值，得到的互信息值也会发生变化，记录其中最大的互信息值为 I(X,xi,xj)。然后执行归一化以比较不同维数下的数据集，并且归一化后的值在 [0,1]之间。通过更改 x，y 的值，可以获得变量之间归一化后的互信息值特征矩阵。特征矩阵的最大值是两个变量之间的最大信息系数 MIC 值。

X={x1,x2,x3,…,xn}样本容量取值为 n，网格化的分数取值小于 B(n)。则最大信息系数可以定义为：

$$M(X)_{x_i, x_j} = \frac{I(X, x_i, x_j)}{\log(\min\{x_i, x_j\})} \quad (5)$$

$$MIC(X) = \max_{xy < B(n)} \{M(X)_{x,y}\} \quad (6)$$

式中，x,y 是在 x 轴 y 轴方向上的划分格子的个数，也就是网格分布，其中 B(n)是一个变量，B(n)的大小一般为数据 n 的 0.6 次方左右，即 B(n)≈n0.6。

## （3） MIC 的计算

在 Python 中我们可以借助 minepy 库来完成 MIC 的计算，本文将 avg_Sentimen 和 avg_Price 作为变量计算二者的 MIC，其结果如图 6 所示：

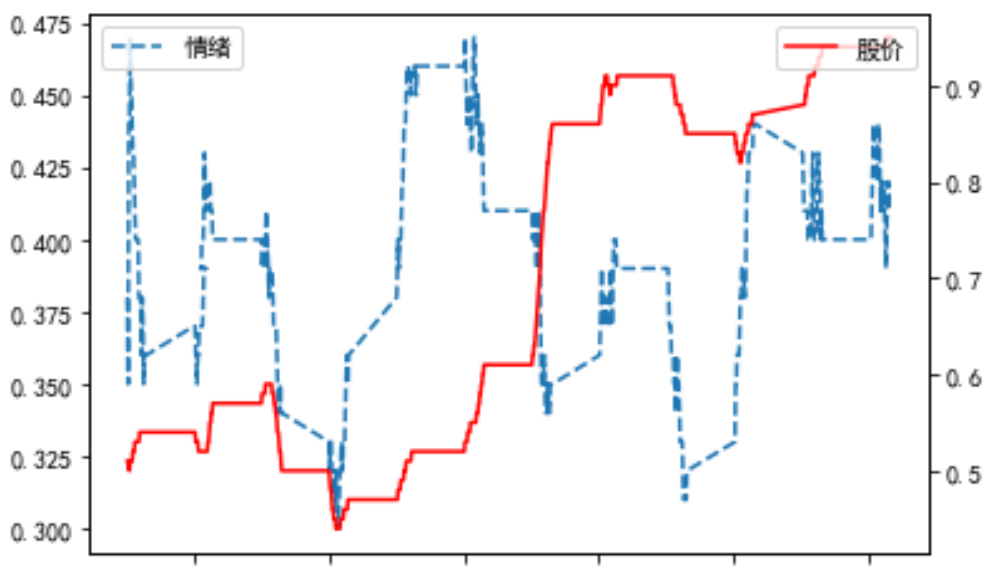

图 6 avg_Sentimen 和 avg_Price 的计算结果

$$MIC（\text{avg\_Sentimen}, \text{avg\_Price}) = 0.3806609398310775$$

MIC 的值约为 0.38，说明 avg_Sentimen 和 avg_Price 呈现一定程度的相关关系。

## 六、 总结

深度学习的不断发展，为我们处理金融问题提供了一个很好的框架，本文也表明了深度学习模型在金融领域的有效性，基于此搭建出来的 BERT 模型很好的解决了 Word2Vec 等深度学习中存在的一词多义的问题，基于金融语料训练出来的 BERT 模型，在股票预测方面有较为良好的表现，在传统方法对股票进行的基础上结合情感分析，能够更好的提高预测的准确率，此外对投资者情绪定量指标的构建以及投资者情绪影响股票市场的研究，有利于投资者在投资中把握走势，依此获得超额收益。另一方面，同时，由于我国股票市场存在大量散户投资者，投资者情绪变动对股票市场的稳定性具有一定的影响，在通过深度学习的方法进一步探究投资者情绪对股票市场的影响机制，将有利于国家监管部门和政策部门对维持股票市场稳定性制定更加合理的政策方针。

# 七、 不足之处与未来展望

对于金融实体的选取，本文选取了股吧作为研究的对象，研究过程中发现，股吧由于其自发性，数据噪声非常大，不利于数据的拟合，会对深度学习模型效果产生影响，后期希望采用财经新闻文本数据进行研究，一方面财经新闻文本数据有着统一的标准，另一方面它能直接反馈上市公司的实际情况，便于与市场拟合。再者，受笔者研究条件、机器性能的限制，深度学习模型在训练过程中不能达到最优的训练效果，后期可以考虑在充足的经费支持下，采用云计算等方式进行研究。

# 参考文献